\documentclass[a4paper]{spie}  %
\usepackage[utf8]{inputenc}

\usepackage{amsmath,amsfonts,amssymb}
\usepackage{graphicx}
\usepackage[colorlinks=true, allcolors=blue]{hyperref}
\usepackage{booktabs}

\title{KalAO the swift adaptive optics imager on 1.2m Euler Swiss telescope in La Silla, Chile}

\author[a]{Janis Hagelberg}
\author[a]{Nathanaël Restori}
\author[a]{François Wildi}
\author[a]{Bruno Chazelas}
\author[b]{Christoph Baranec}
\author[c,d,e]{Olivier Guyon}
\author[a]{Ludovic Genolet}
\author[a]{Michaël Sordet}
\author[f]{Reed Riddle}
\affil[a]{Geneva Observatory, University of Geneva, Geneva, Switzerland}
\affil[b]{Institute for Astronomy, University of Hawai'i at Mānoa, Hilo, HI, USA}
\affil[c]{Astrobiology Center, National Institutes of Natural Sciences, Tokyo, Japan}
\affil[d]{Steward Observatory, The University of Arizona, Tucson, USA}
\affil[e]{Subaru Telescope, National Astronomical Observatory of Japan, Hilo, HI, USA}
\affil[f]{Caltech Optical Observatories, California Institute of Technology, Pasadena, CA, USA}

\authorinfo{Further author information: (Send correspondence to J.H.)\\J.H.: E-mail: janis.hagelberg@unige.ch}

\begin{document} 
\maketitle

\begin{abstract}
KalAO is a natural guide star adaptive optics (AO) imager to be installed on the second Nasmyth focus of the 1.2m Euler Swiss
 telescope in La Silla, Chile. The initial design of the system is inspired on RoboAO with modifications in order to
  operate in natural guide star (NGS) mode.
KalAO was built to search for binarity in planet hosting stars by following-up candidates primarily from the TESS satellite survey. The optical design is optimised for the 450-900 nm wavelength 
range and is fitted with SDSS \emph{g,r,i,z} filters.
The system is designed for wavefront control down to $I$-magnitude 11 stars in order to probe the same parameter
 space as radial velocity instruments such as HARPS and NIRPS.
The principal components of the system are an 11x11 10.9 cm sub-apertures Electron Multiplying CCD (EMCCD) 
Shack-Hartmann wavefront sensor, 
a 140 actuators Microelectromechanical systems (MEMS) deformable mirror, a fast tip/tilt mirror, and a graphics processing unit (GPU) powered glycol cooled real-time computer.
It is designed to run at up to 1.8kHz in order to detect companions as close as the 150mas visible-light
 diffraction limit. The real-time adaptive optics control is using the CACAO software running on GPUs.
The instrument is planned for commissioning early 2021 in Chile if the covid restrictions are lifted.
\end{abstract}


\section{INTRODUCTION}
\label{sec:intro}  

KalAO is a new high-resolution high-cadence imager which will be installed on the 1.2m Swiss
telescope Euler in Chile in early 2021. The primary goal of KalAO is to observe stars where a planet transit has been detected by the TESS satellite mission in order to search for additional stellar companions. 
The KalAO survey will target candidate and confirmed transiting planets to verify if they are located within a binary, with
80 nights on the Swiss telescope dedicated to this project over the next 2.5 years.  
It will also play a planet candidate vetting role in the TESS follow-up observations effort and possibly other surveys such as PLATO.

\begin{figure} [ht]
	\begin{center}
		\begin{tabular}{c} %
			\includegraphics[width=0.9\linewidth]{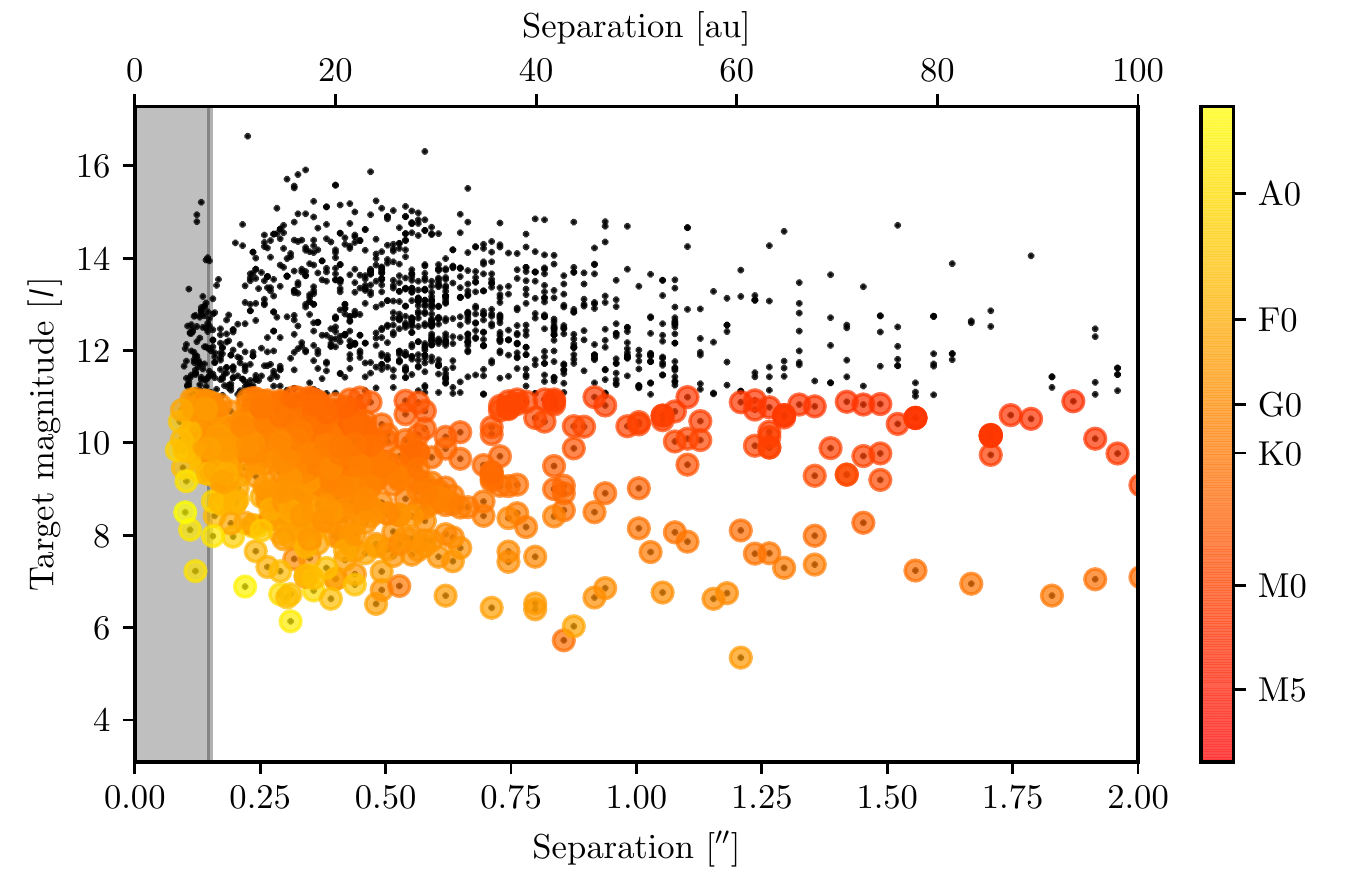}
		\end{tabular}
	\end{center}
	\caption[example] 
	{ \label{fig:sulli} 
		Simulated TESS stellar contaminants yield, based on the sample from Ref.~\citenum{sullivan_transiting_2015}. 
		The greyed area represent the inaccessible parameter space due to the diffraction limit at 0.15". 
		Black dots are contaminants around stars fainter than $I$ mag 11 and thus too faint for good performance adaptive optics and radial velocity.}
\end{figure}

Using the detection yield of the TESS space mission\cite{ricker_transiting_2014} simulated by \citenum{sullivan_transiting_2015}, we derived the expected yield of stellar companions to TESS planet candidates (Fig. \ref{fig:sulli}). 
To combine the detections with radial velocity characterisation (mainly using Coralie, HARPS, or NIRPS), 
the targets in the sample need to be bright enough. 
Stars fainter than magnitude 11 in $I$ become more challenging for radial velocity follow-up. Binary stars with
a separation smaller than 0.1" are expected to yield a strong radial velocity signal typical of short period binary stars.
To reach such small separations on a 1.2m telescope we thus have to observe in the visible.
These considerations on the science case thus result the following top requirements for KalAO:

\begin{itemize}
    \item AO down to $I$ magnitude 11
    \item Wavelength range: 450--900 nm
    \item Diffraction limited FWHM: 0.11" @ 635 nm 
    \item Minimal field of view: 10"
    \item $>15$\% Strehl in $z$ band
\end{itemize}

\section{DESCRIPTION OF KALAO}
\label{sec:description}

The design of the system was inspired on RoboAO\cite{2014ApJ...790L...8B, 2018SPIE10703E..27B} with the major modification being that KalAO
will only operate in natural guide star (NGS) mode.

KalAO will be mounted on the second Nasmyth focus of the Swiss 1.2m Euler telescope located at ESO La Silla Observatory in Chile. The optical bench 
will be attached vertically on the telescope and will rotate along the axis of elevation. The field of view will thus rotate while the telescope
pupil will remain fixed. This will make it possible to apply point spread function (PSF) subtraction methods such as angular differential imaging\cite{marois_angular_2006} (ADI).

\begin{figure} [ht]
	\begin{center}
		\begin{tabular}{c} %
			\includegraphics[width=0.6\linewidth]{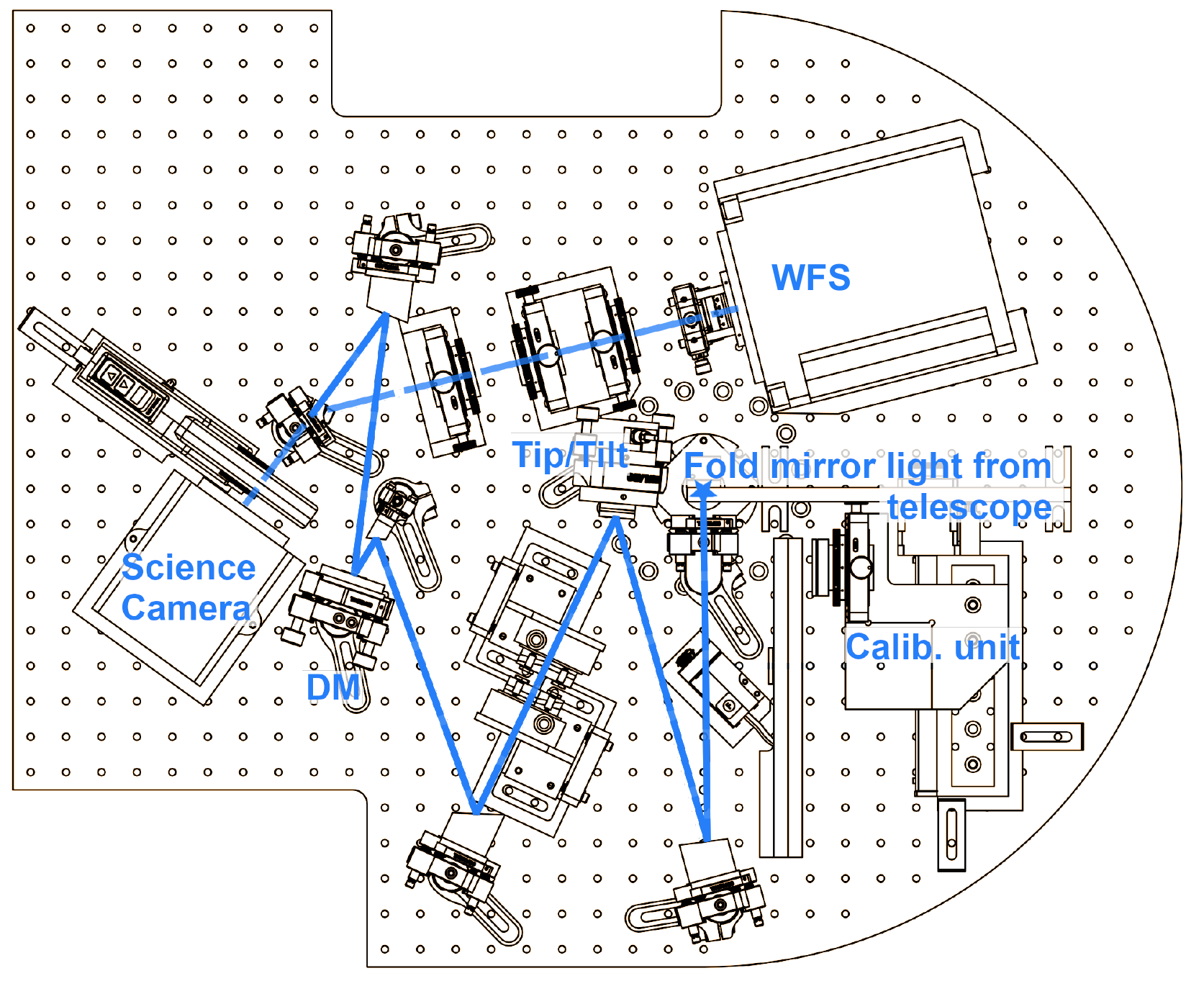}
		\end{tabular}
	\end{center}
	\caption[example] 
	{ \label{fig:zemax} 
		Beam path in sky observation mode. The light starts by coming through the bench from below at the position of the
		blue star.}
\end{figure} 

\begin{figure} [ht]
	\begin{center}
		\begin{tabular}{c} %
			\includegraphics[width=0.97\linewidth]{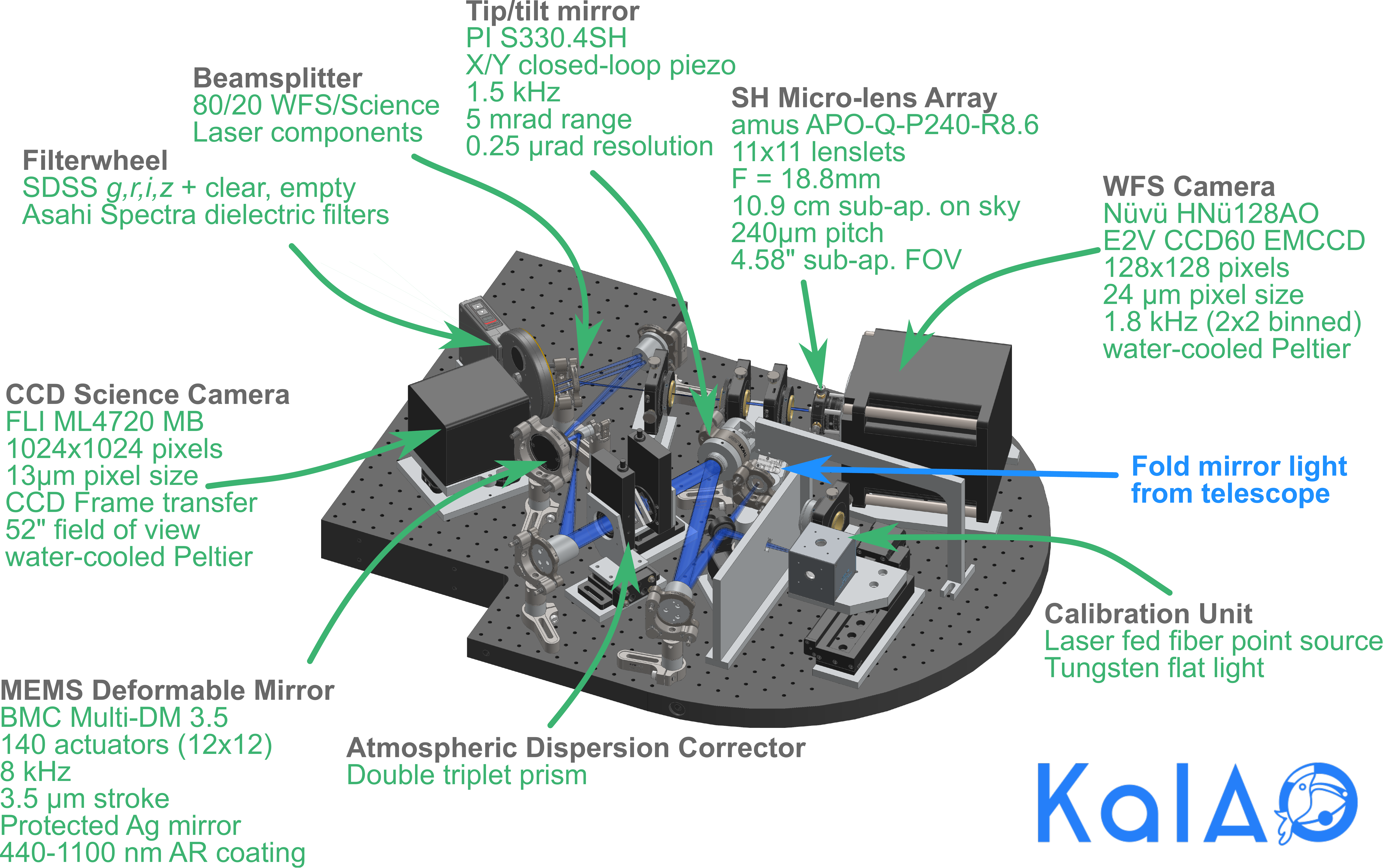}
		\end{tabular}
	\end{center}
	\caption[example] 
	{ \label{fig:3d} 
Three dimensional view of KalAO with the description of the main components. The light path is illustrated in blue, starting in the centre of the bench with the light from the telescope coming from beneath the bench.}
\end{figure}

The optical design (given on figure \ref{fig:zemax}) consists of a shutter, used to close the instrument when not in use or during internal calibrations, and a fold mirror, used to fold the light beam from the telescope inside a plane parallel to the instrument base plate, at the entrance of the instrument. Off-axis parabolic mirrors with protected silver coating relay the pupil from the telescope on a Physik Instrument S330.4SH tip-tilt mirror followed by a Boston Micromachine Corp Multi-DM 3.5 deformable mirror and to finally image the sky on a FLI ML4720 MB CCD camera. An atmospheric dispersion corrector (ADC) is placed in the collimated beam after the tip-tilt mirror. A beam splitter splits the light between the science camera and the Shack-Hartmann wavefront sensor with a 20\%/80\% ratio. A Thorlabs 6-position filter wheel, with SDSS \emph{g,r,i,z} interference filters from Asahi Spectra and a clear window, is placed in front of the science camera. The wavefront sensor consists of a Nüvü HNü128AO EMCCD camera and a microlens array from a\textmu{}s microptics. A Calibration Unit with a laser-fed fibre and a tungsten-fed integrating sphere on a translation stage is used to do calibration when the instrument is not on the sky. A flip mirror allows to switch between the Calibration Unit and the telescope. The main specifications of the instrument components is
listed in table \ref{tab:kalao_specs}, while their position on the bench is illustrated on figure \ref{fig:3d}.

\begin{table}[ht]
\caption{Specifications of the main instrument components.} 
\label{tab:kalao_specs}
\begin{center}       
\begin{tabular}{ l l } 
\toprule
Tip/tilt mirror &  PI S330.4SH \\
                &  X/Y closed loop piezo stage\\
                & 1.5 kHz resonant frequency with mirror\\
                & 5 mrad range\\
                & 0.25 $\mu$rad resolution\\
\midrule
Deformable mirror & BMC Multi-3.5\\
                & MEMS technology\\
                & 140 actuators (12x12)\\
                & 8 kHz max frequency \\
                & 3.5 $\mu$m stroke\\
                & Protected silver mirror coating\\
                & 440-1100 nm anti-reflection coating on entry window\\\midrule
Beam-splitter    & Laser components dielectric coating\\
                & 80\% light reflected to wavefront sensor\\
                & 20\% transmitted to science camera\\\midrule
Filter-wheel     & Thorlabs six positions FW102C\\
                & SDSS $g$, $r$, $i$, $z$ dielectric filters (Asahi Spectra)\\
                & 1 clear window\\
                & 1 empty slot\\
                \midrule
Science camera  & FLI ML4720 MB\\
                & CCD frame transfer technology\\
                & 1024x1204 pixels\\
                & 13 $\mu$m pixel size\\
                & 52" field of view\\
                & Peltier CCD cooler linked to external water cooling\\
                \midrule
Shack-Hartmann micro-lens array & a\textmu{}s microptics APO-Q-P240-R8.6\\
                & 11x11 lenslets \\
                & F = 18.8mm\\
                & 10.9 cm sub-apertures on sky\\
                & 4.58" sky field of view per sub-aperture\\
                & 240 $\mu$m pitch\\\midrule
Wavefront sensor camera & Nüvü HNü128AO\\
                & E2V CCD60 EMCCD\\
                & 128x128 pixels, binned at 64x64\\
                & 24 $\mu$m pixel size\\
                & 1.8 kHz (2x1) binned \\
                & Peltier CCD cooler linked to external water cooling\\\midrule
Real-time computer & 8 core Intel i9-9900K (water cooled)\\
                & Asus WS Z390 PRO mainboard\\
                & ZOTAC GeForce RTX 2080Ti (ZT-T20810K-30P) GPU w-cooled\\
                & HyperX Predator DDR4 64GB Kit (4 x 16GB) 2666MHz\\
\bottomrule
\end{tabular}
\end{center}
\end{table}

The instrument is built around a base plate made out of machined aluminium. All the optics are held inside commercial off-the-shelf Thorlabs polaris mounts, such as kinematic mounts, and translation and rotation stages. Aluminium blocks and pillars were machined at the mechanical workshop of the Geneva Observatory to hold the mounts at the correct height.

A flexible interface made out of stainless steel and aluminium is used to mount the instrument on the telescope and to accommodate the difference of the telescope (steel) and the base plate (aluminium) the thermal expansion. A fibreglass reinforced composite hood is used to protect the instrument from the environment and from stray light while keeping the 
weight low.

\begin{figure} [ht]
	\begin{center}
		\begin{tabular}{c c} %
			\includegraphics[width=0.46\linewidth]{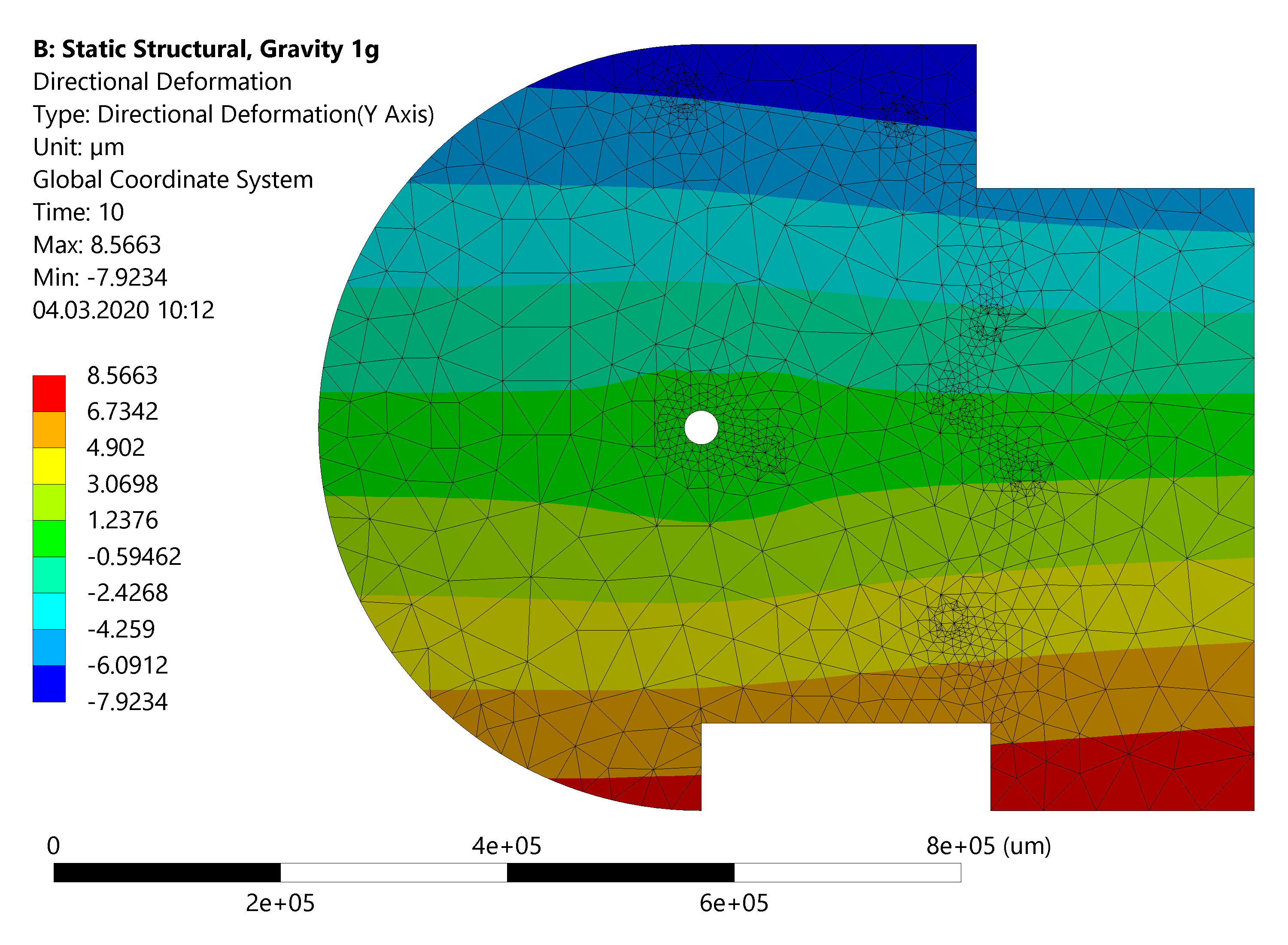}
&
			\includegraphics[width=0.46\linewidth]{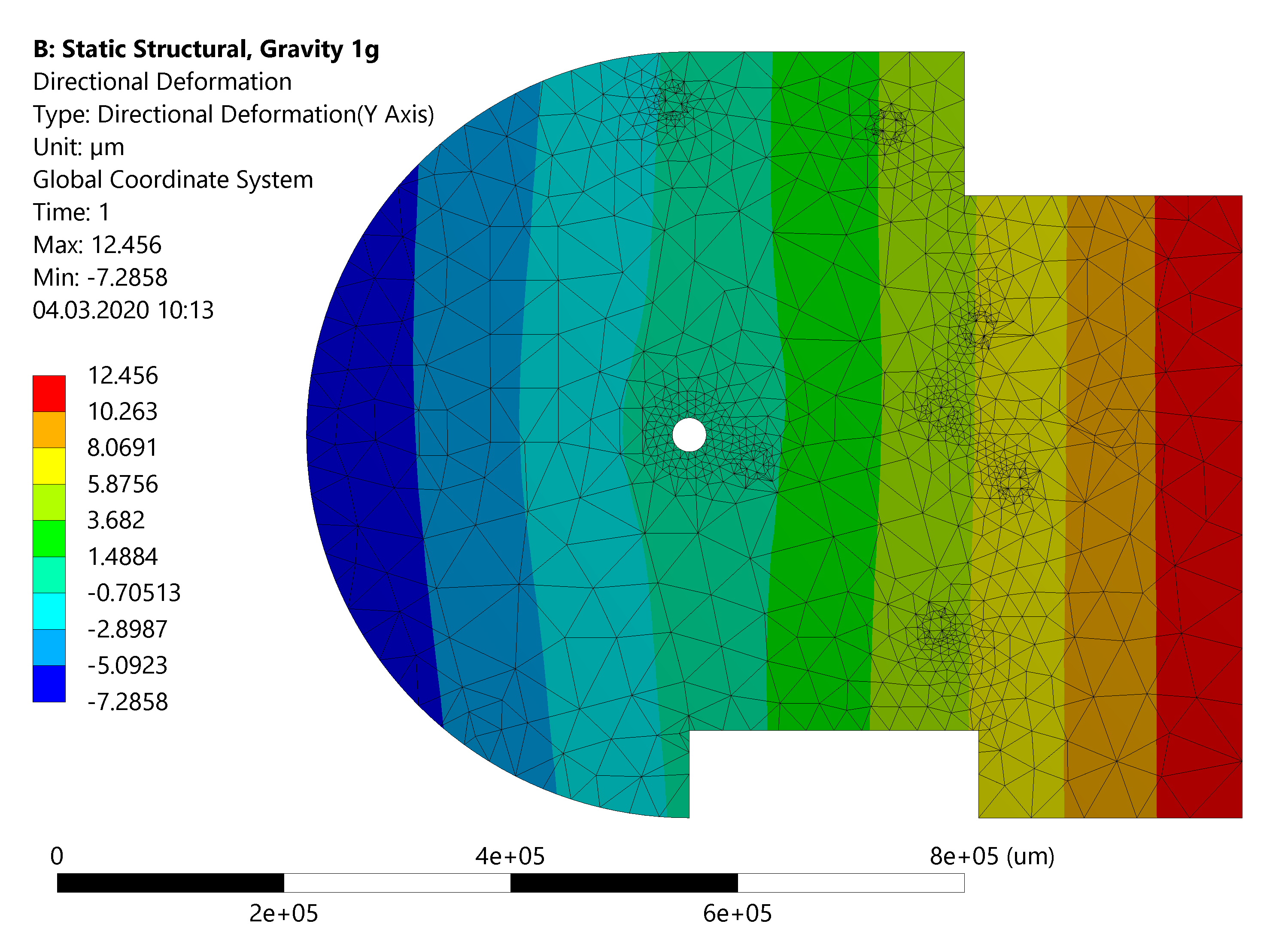}\\
			\footnotesize{ a) Gravity deformation at elevation $0^\circ$ }&\small{ b) Gravity deformation at elevation $90^\circ$}\\
			\includegraphics[width=0.46\linewidth]{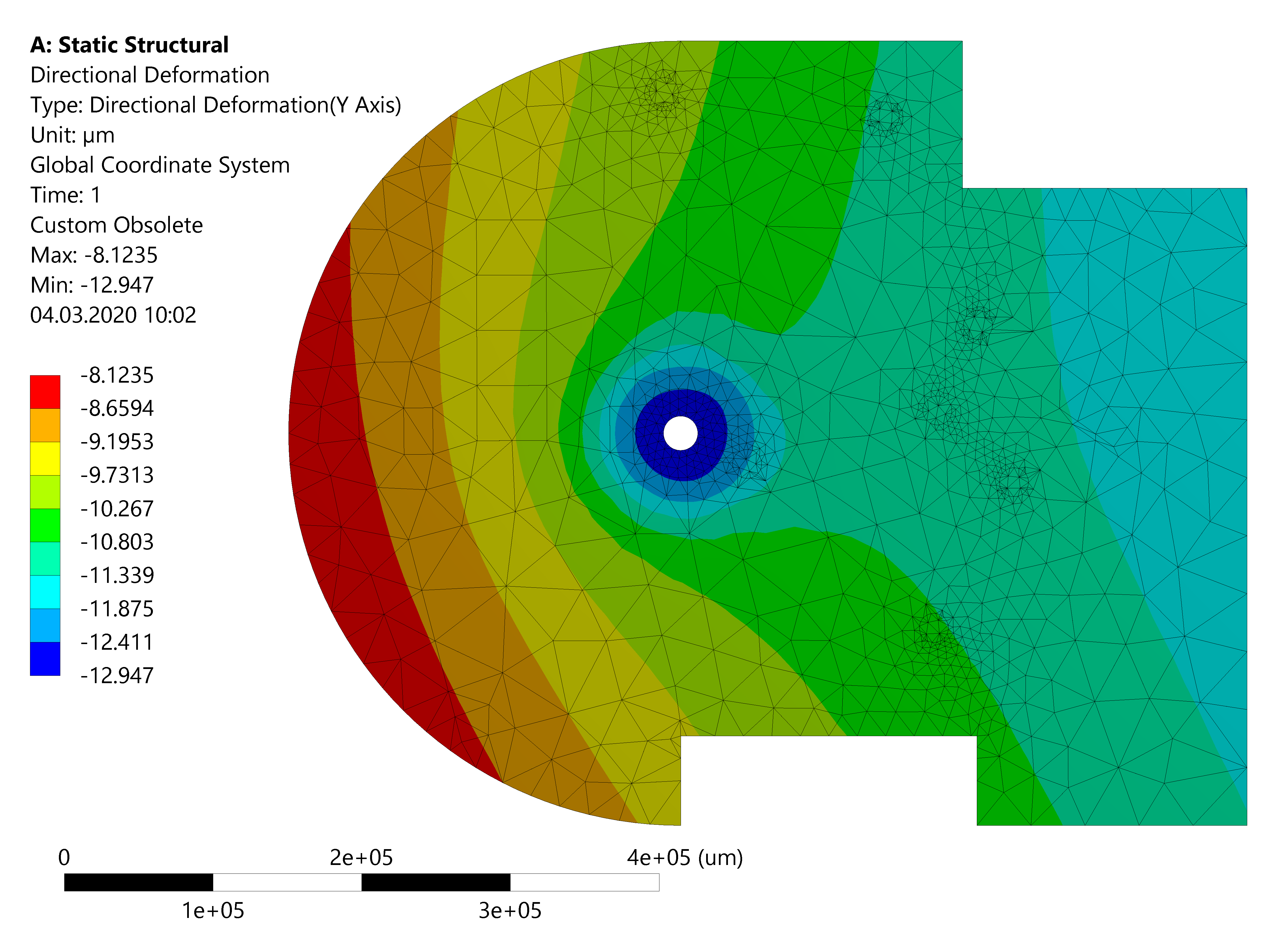}&
			\includegraphics[width=0.46\linewidth]{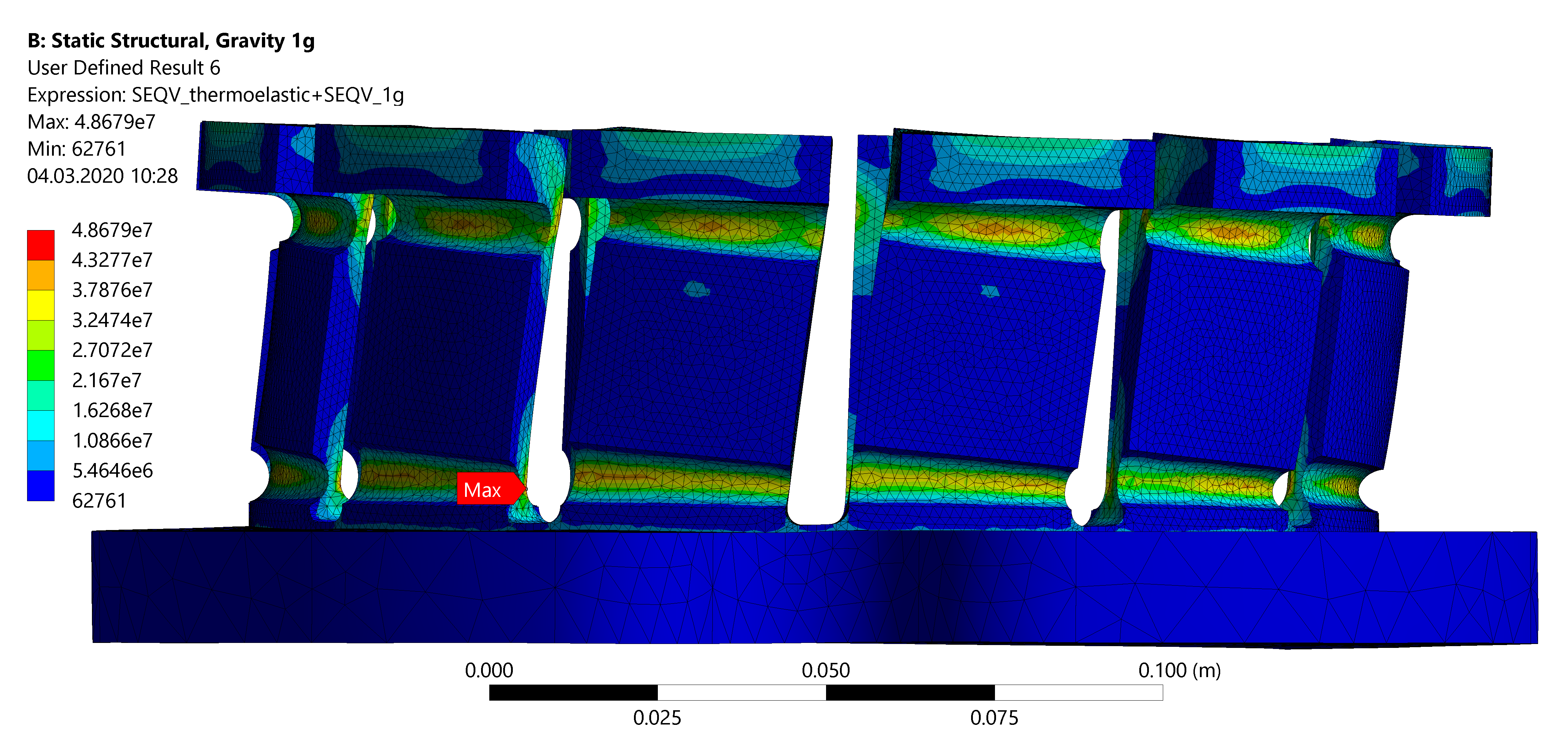}\\
			\footnotesize{c) Out of plane thermal deformation }&\small{d) Stress on the instrument--telescope interface}
		\end{tabular}
	\end{center}
	\caption[example] 
	{ \label{fig:ansys} 
	Finite element modelling analysis to validate the optical bench stability caused by gravity (\emph{a, b, d}) and thermal
	gradient (\emph{c, d}) when installed vertically on the telescope	with all components present.}
\end{figure}

As the instrument will be fixed on the telescope and moving along the elevation axis we carried out a mechanical study
to ensure that our solution is stiff enough to minimise non-common path aberration (NCPA) while being at the same time
light enough to be held by the telescope.
Two type of loading were modelled: a thermal loading of $-\Delta 25^\circ$C to simulate the cooling of the
instrument during the night, and a static loading, to simulate the deformation of the instrument due
to its own weight for different elevation of the telescope (Fig. \ref{fig:ansys}).

\begin{figure} [ht]
	\begin{center}
		\begin{tabular}{c} %
			\includegraphics[height=0.65\linewidth, angle=90]{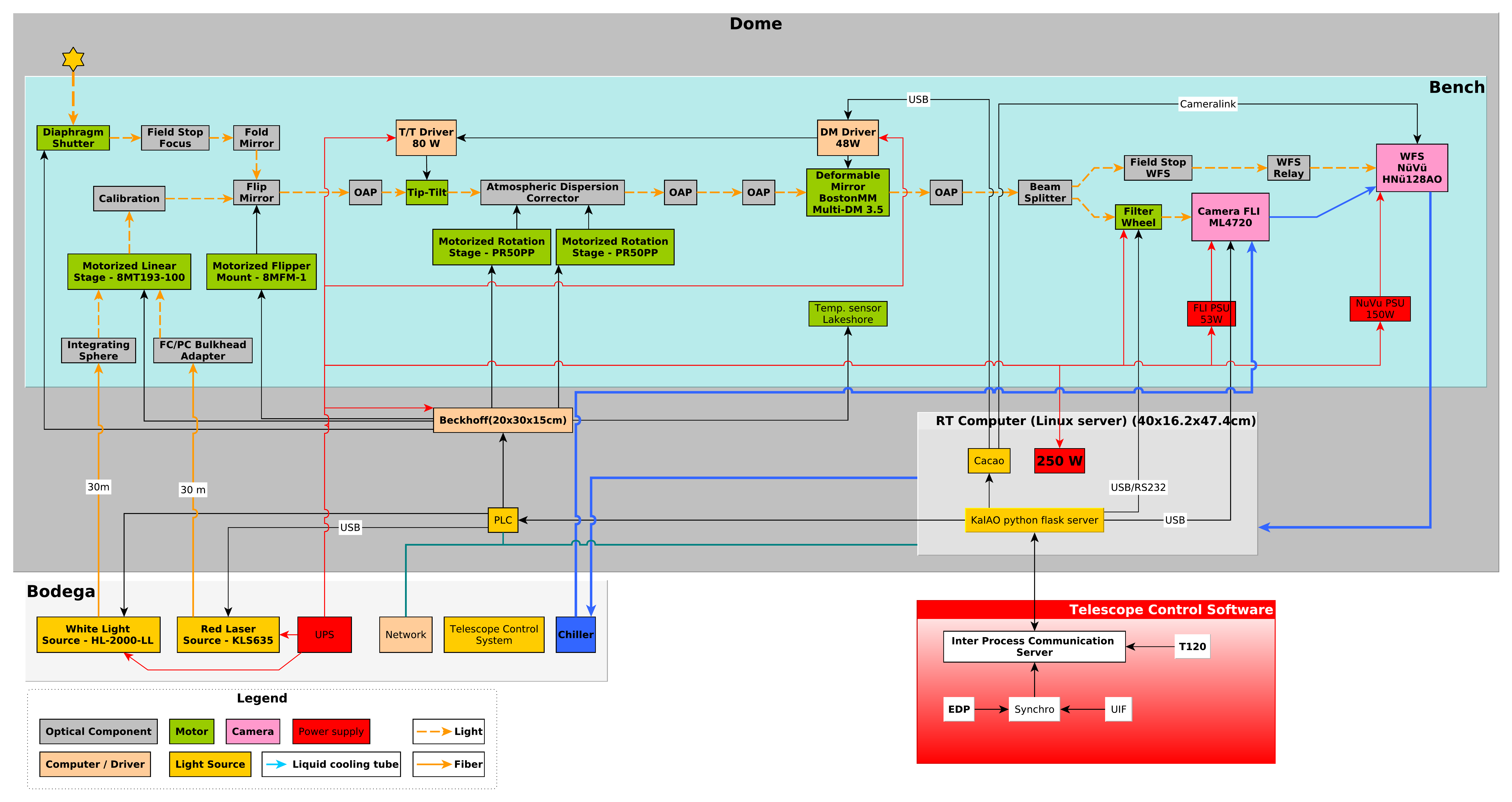}
		\end{tabular}
	\end{center}
	\caption[example] 
	{ \label{fig:diagram} 
		System diagram of KalAO and its integration within the Swiss 1.2m Euler telescope infrastructure.}
\end{figure}

The real time computer (RTC) which is driving the AO loop and the science camera will be situated within the telescope dome.
To minimise dome seeing from the RTC thermal emission the whole system is glycol-cooled. The main components of the RTC are an
an 8 core Intel i9-9900K on an Asus WS Z390 PRO mainboard along with a ZOTAC GeForce RTX 2080Ti (ZT-T20810K-30P) GPU card.
The RTC assembly is encased in a Silverstone SST-CS350B chassis with a Hydro PTM water-cooled power supply unit 
(Fig. \ref{fig:rtc_photo}). 
Besides the adaptive optics loop, the RTC also controls all the other components of KalAO as is illustrated in the system 
diagram \ref{fig:diagram}.

The adaptive optics control system relies on CACAO\cite{guyon_compute_2018} and can run on at up to 1.8kHz, which is the
wavefront sensor (WFS) camera maximum frame rate. Advanced AO computation such as predictive control is running on the GPU for best performance.
Custom modules were built for the interfacing with the deformable mirror and the EMCCD camera. A Shack-Hartmann module was also implemented
in CACAO to compute the slopes of the wavefront which previously was only running on pyramid based systems. 

\section{STATUS}
\label{sec:status}

\begin{figure} [ht]
	\begin{center}
		\begin{tabular}{c c} %
			\includegraphics[height=0.24\textheight]{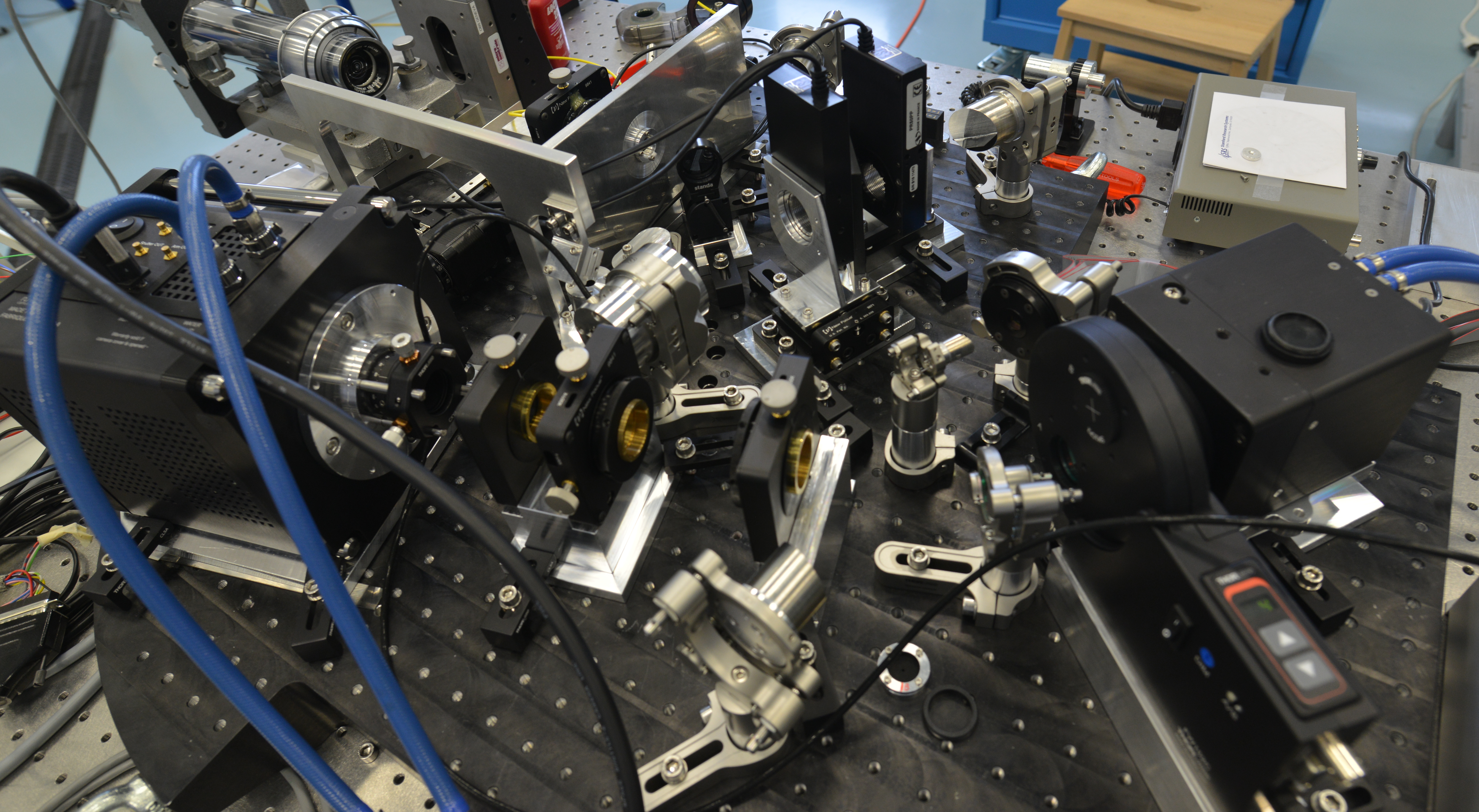}&
			\includegraphics[height=0.24\textheight]{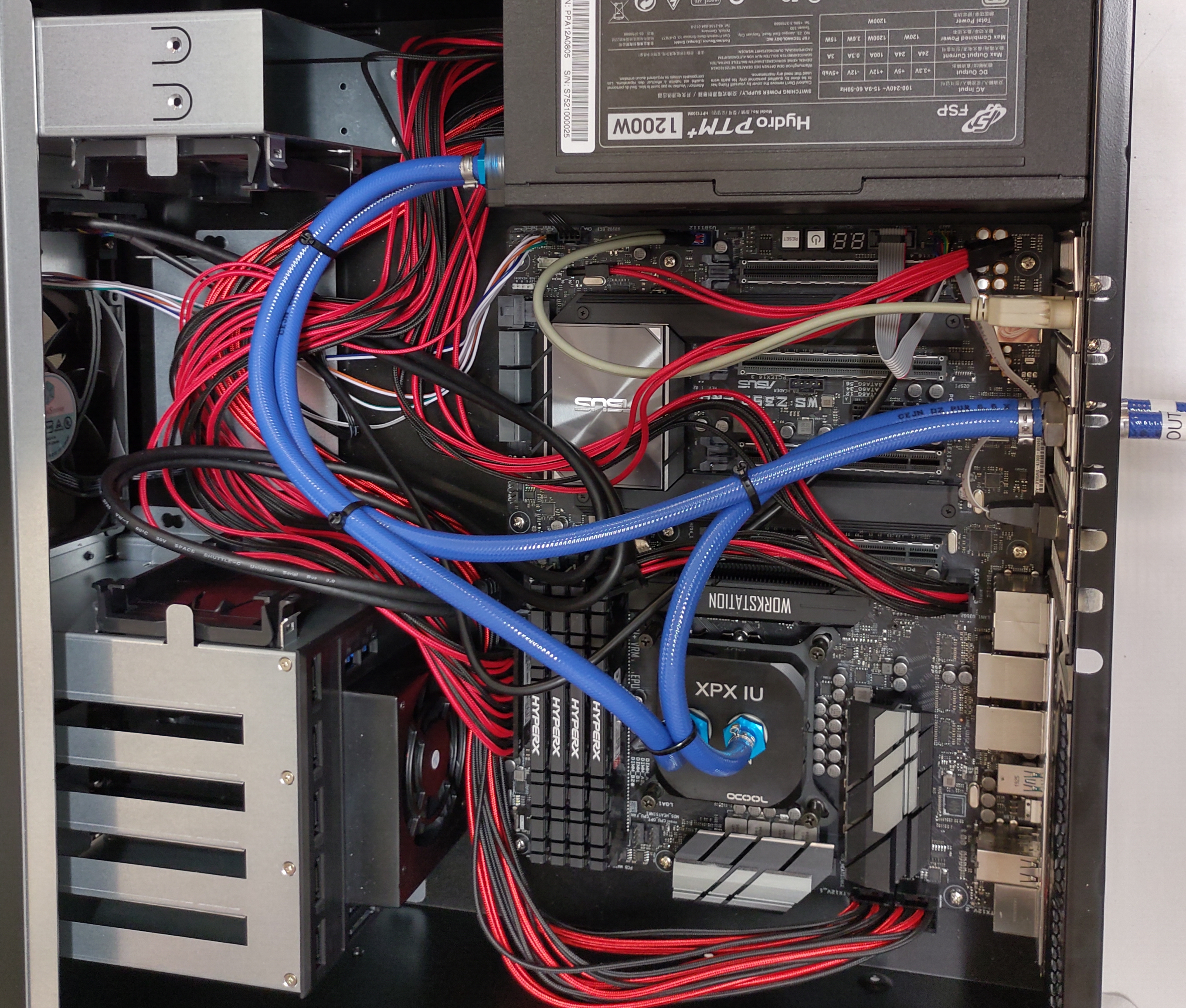}\\
			a) RTC & b) Bench
		\end{tabular}
	\end{center}
	\caption[example] 
	{View of the KalAO bench while testing adaptive optics loop closure in lab and of the KalAO RTC before the GPU was added. The blue tubes are the glycol cooling system.\label{fig:bench_photo}\label{fig:rtc_photo} }
\end{figure}

Hardware assembly of the optical bench is completed (see figure \ref{fig:bench_photo}). The adaptive optics loop has also been closed in lab with a simple atmosphere 
turbulence simulator at 1.8kHz speed with two frames of latency. Further performance assessments are planned with a properly characterised simulator.
Software integration of the instrument control system with the telescope control system is still ongoing but is soon to be finalised.
As part of the integration on the telescope the third mirror of the telescope needs to be upgraded in order to be able to switch
between the two Nasmyth and the Cassegrain foci.
Even though KalAO is almost ready to start operation, due to covid health restrictions a date for the commissioning run in Chile could not yet be determined.

\section{FUTURE WORK}
\label{sec:future}

KalAO will be used as a prototype platform for RISTRETTO\cite{lovis_atmospheric_2017, chazelas_ristretto_2020}.
The design is such that it can be reconfigured in order to host the
RISTRETTO spectrograph fibre injection module. The pierced mirror with the integral field unit (IFU) would be placed at the position of 
the science camera, while the camera will be re-positioned in order to be used as a guiding and fibre centring camera.
This temporary setup will be used for a few months in order to validate on sky the RISTRETTO concept.

\section*{ACKNOWLEDGMENTS}

This project is supported by SNSF through the Ambizione grant \#PZ00P2\_180098 (PI. Janis Hagelberg).

\bibliography{KalAO_SPIE, manual_bib} %
\bibliographystyle{spiebib} %

\end{document}